\documentclass[12pt]{article}
\usepackage{epsfig}

\begin{document}

\title{Polynomial Solutions of Shcr\"odinger Equation with the Generalized Woods$-$Saxon Potential}
\author{C\"uneyt Berkdemir $^a$, Ay\c se Berkdemir $^a$ and  Ramazan Sever $^{b}$$\thanks{Corresponding author:
sever@metu.edu.tr}$\\
{\small {\sl $^a$ Department of Physics, Faculty of Arts and
Sciences, Erciyes University,}} {\small {\sl 38039, Kayseri, Turkey}}\\
{\small {\sl $^b$ Department of Physics, Middle East Technical
University,}} {\small {\sl 06531, Ankara, Turkey}}}

\date{\today}

\maketitle

\begin{abstract}

The bound state energy eigenvalues and the corresponding
eigenfunctions of the generalized Woods$-$Saxon potential are
obtained in terms of the Jacobi polynomials. Nikiforov-Uvarov
method is used in the calculations. It is shown that the results
are in a good agreement with the ones obtained before.

\end{abstract}

\baselineskip=22pt plus 1pt minus 1pt

\vspace{0.5cm}



\newpage

\section{Introduction}

Exact solution of Schr\"odinger equation for central potentials
has generated much interes in recent years. So far, these
potentials are the parabolic type potential \cite {1}, the Eckart
potential \cite {1, 2, 3}, the Fermi-step potential \cite {2, 3},
the Rosen- Morse potential \cite {4}, the Ginocchio barrier \cite
{5}, the Scarf barrier \cite {6}, the Morse potential \cite {7}
and a potential which interpolates between Morse and Eckart
barriers \cite {8}. In addition, many authors have studied on
exponential type potentials \cite {9, 10, 11, 12, 13} and quasi
exactly solvable quadratic potentials \cite {14, 15, 16}. The
exact solutions for these models have been obtained analytically.

Recently, an alternative method known as the Nikiforov-Uvarov (NU)
method has been introduced for solving the Schr\"{o}dinger
equation (SE). There have been several applications of SE with
some well-known potentials \cite {17, 18, 19}, Dirac, Klein-Gordon
and Duffin-Kemmer-Petiau equations for a Coulomb type potential by
using this method as well \cite {21, 22, 23}.

In the present work, the real-valued bound-state energies of the
generalized Woods-Saxon potential are evaluated through the NU
Method \cite {24} by following the framework of quantum mechanics.
This method is based on solving the time-independent Schr\"odinger
equation by reduction to a generalized equation of hypergeometric
type. Energy eigenvalues and the corresponding eigenfuctions are
calculated exactly. The generalized Wood-Saxon potential as a
shell model is selected. It can be used for describing metallic
clusters in a successful way and for lighting the central part of
the interaction neutron with one heavy nucleus \cite {25, 26}.

This paper is arranged as follows: In Sec. II we introduce the NU
method. In Sec. III we apply the method to solve the
Schr\"{o}dinger equation with the generalized Woods-Saxon
potential. Shapes of the potential and energy are studied in Sec.
IV. In Sec. V we discuss the results.

\section{Nikiforov-Uvarov Method}

The NU method provides us an exact solution of non$-$relativistic
Schr\"odinger equation for certain kind of potentials \cite {24}.
The method is based on the solutions of general second order
linear differential equation with special orthogonal functions
\cite {27}. For a given real or complex potential, the
Schr\"odinger equation in one dimension is reduced to a
generalized equation of hypergeometric type with an appropriate
$~s=s(x)~$ coordinate transformation. Thus it can be written in
the following form,
\begin{equation}
\label{eq1} \psi ^{\prime \prime }(s)+\frac{\stackrel{\sim }{\tau
}(s)}{\sigma }\psi ^{\prime }(s)+\frac{\stackrel{\sim }{\sigma
}(s)}{\sigma ^{2}(s)}\psi (s)=0
\end{equation}
where $~\sigma (s)$ and $~\stackrel{\sim }{\sigma }(s)~$ are
polynomials, at most second$-$degree, and $~\stackrel{\sim }{\tau
}(s)~$ is a first$-$degree polynomial. To find a particular
solution of Eq.(\ref{eq1}) by separation of variables, we use the
following the transformation
\begin{equation}
\label{eq2} \psi(s)=\phi(s)y(s)
\end{equation}
This reduces Schr\"odinger equation, Eq.(1), to an equation of
hypergeometric type,
\begin{equation}
\label{eq3} \sigma(s)y^{\prime \prime }+\tau(s)y^{\prime }+\lambda
y=0,
\end{equation}
where $\phi(s)$ satisfies $~\phi (s)^{\prime }/\phi (s)=\pi
(s)/\sigma (s)$. $y(s)~$ is the hypergeometric type function whose
polynomial solutions are given by Rodrigues relation
\begin{equation}
\label{eq4}
y_{n}(s)=\frac{B_{n}}{\rho (s)}\frac{d^{n}}{ds^{n}}\left[ \sigma ^{n}(s)\rho (s)%
\right] ,
\end{equation}
where $B_{n}$ is a normalizing constant and the weight function
$\rho $ must satisfy the condition \cite {24}
\begin{equation}
\label{eq5} (\sigma \rho )^{\prime }=\tau \rho.
\end{equation}
The function$~\pi~$ and the parameter$~\lambda~$ required for this
method are defined as
\begin{equation}
\label{eq6} \pi =\frac{\sigma ^{\prime }-\stackrel{\sim }{\tau
}}{2}\pm
\sqrt{\left(\frac{\sigma^{\prime}-\stackrel{\sim}{\tau}}{2}\right)^{2}-\stackrel{\sim}{\sigma}+k{\sigma}}
\end{equation}
and
\begin{equation}
\label{eq7} \lambda =k+\pi ^{\prime }.
\end{equation}
Here, $\pi(s)$ is a polynomial with the parameter $s$ and the
determination of $k$ is the essential point in the calculation of
$\pi(s)$. Thus, in order to find the value of $k$, the expression
under the square root must be square of a polynomial. Hence, a new
eigenvalue equation for the Schr\"{o}dinger equation becomes
\begin{equation}
\label{eq8} \lambda =\lambda _{n}=-n\tau ^{\prime
}-\frac{n(n-1)}{2}\sigma ^{\prime \prime },~~~(n=0,1,2,...)
\end{equation}
where
\begin{equation}
\label{eq9} \tau (s)=\stackrel{\sim }{\tau }(s)+2\pi (s),
\end{equation}
and it will have a negative derivative.

\section{Generalized Woods$-$Saxon Potential}

The interactions between nuclei are commonly described by using a
potential that consist of the Coulomb and the nuclear potentials.
These potentials are usually taken to be of the Woods-Saxon form.
As an example it can be given as the generalized Woods-Saxon
potential \cite {28}
\begin{equation}
\label{eq10}
V(r)=-\frac{V_{0}}{1+e^{\left(\frac{r-R_{0}}{a}\right)}}-\frac{C.e^{\left(\frac{r-R_{0}}{a}\right)}}{\left(1+e^{\left(\frac{r-R_{0}}{a}\right)}\right)^2}~,
\end{equation}
where $V_0$ is the potential depth, $R_{0}$ is the width of the
potential and $a$ is the surface thickness which is usually
adjusted to the experimental values of ionization energies. In
order to calculate the energy eigenvalues and the corresponding
eigenfunctions, the potential function given by Eq.(\ref{eq10}) is
substituted into the radial part of Schr\"odinger equation:
\begin{equation}
\label{eq11}
\psi^{\prime\prime}(r)+\frac{2m}{\hbar^2}\left[E+\frac{V_{0}}{1+qe^{2\alpha
r}}+\frac{Ce^{2\alpha r}}{\left(1+qe^{2\alpha
r}\right)^2}\right]\psi(r)=0.
\end{equation}
Here, some assignments are made in the radial Schr\"odinger
equation such as $R(r)=\psi(r)/r$, $r-R_0\equiv r$ and $1/a\equiv
2\alpha$. In addition, \textit {q} is deformation parameter and an
arbitrary real constant within the potential.

Now, in order to apply the NU$-$method, we rewrite Eq.(\ref{eq11})
by using a new variable of the form $s=-e^{2\alpha r}$,
\begin{equation}
\label{eq12}
\frac{d^2\psi(s)}{ds^2}+\frac{1}{s}\frac{d\psi(s)}{ds}+\frac{m}{2\hbar^2\alpha^2s^2}\left[E+\frac{V_{0}}{(1-qs)}-\frac{Cs}{\left(1-qs\right)^2}\right]\psi(s)=0.
\end{equation}
By introducing the following dimensional parameters
\begin{equation}
\label{eq13}
\varepsilon=-\frac{mE}{2\hbar^2\alpha^2}>0~~~(E<0),~~~~~~\beta=\frac{mV_0}{2\hbar^2\alpha^2}~~~(\beta>0),~~~~~~\gamma=\frac{mC}{2\hbar^2\alpha^2}~~~(\gamma>0)
\end{equation}
which leads to a hypergeometric type equation defined in
Eq.(\ref{eq1}):
\begin{equation}
\label{eq14}
\frac{d^2\psi(s)}{ds^2}+\frac{1-qs}{s(1-qs)}\frac{d\psi(s)}{ds}+\frac{1}{s^2(1-qs)^2}\times\left[-\varepsilon
q^2 s^2+(2\varepsilon q-\beta
q-\gamma)s+\beta-\varepsilon\right]\psi(s)=0.
\end{equation}
After the comparison of Eq.(\ref{eq14}) with Eq.(\ref{eq1}), we
obtain the corresponding polynomials as
\begin{equation}
\label{eq15} \stackrel{\sim}{\tau }(s)=1-qs,~~~{\sigma
}(s)=s(1-qs),~~~\stackrel{\sim}{\sigma }(s)=-\varepsilon q^2
s^2+(2\varepsilon q-\beta q-\gamma)s+\beta-\varepsilon.
\end{equation}
Substituting these polynomials into Eq.(\ref{eq6}), we obtain
$\pi$ function as
\begin{equation}
\label{eq16} \pi (s)=-\frac{qs}{2}\pm
\frac{1}{2}\sqrt{\left(q^2+4\varepsilon
q^2-4kq\right)s^2+4\left(\beta q+\gamma-2\varepsilon
q+k\right)s+4\left(\varepsilon-\beta\right)}
\end{equation}
taking $ \sigma ^{\prime }(s)=1-2qs$. The discriminant of the
upper expression under the square root has to be zero. Hence, the
expression becomes the square of a polynomial of first degree;
\begin{equation}
\label{eq17} \left(q^2+4\varepsilon q^2-4kq\right)s^2+4\left(\beta
q+\gamma-2\varepsilon
q+k\right)s+4\left(\varepsilon-\beta\right)=0,
\end{equation}
and
\begin{equation}
\label{eq18} \Delta=\left[4\left(\beta q+\gamma-2\varepsilon
q+k\right)\right]^2-4\times4\left(\varepsilon-\beta\right)\left(q^2+4\varepsilon
q^2-4kq\right)=0.
\end{equation}
When the required arrangements are done with respect to the
constant $k$, its double roots are derived as $k_{1,2}=(\beta
q-\gamma)\pm
q\sqrt{\left(\varepsilon-\beta\right)(1+\frac{4\gamma}{q})}$.

Thus substituting, these values for each \textit{k}~into
Eq.(\ref{eq16}) following possible solution is obtained for $\pi
(s)$

\begin{equation}
\label{eq19} \pi(s) = -\frac{qs}{2}\pm
\frac{1}{2}\left\{\begin{array}{ccc}
\left[(2\sqrt{\varepsilon-\beta}-\sqrt{1+\frac{4\gamma}{q}}~)qs-2\sqrt{\varepsilon-\beta}~\right],
& \hskip 0.5cm \mbox{for} \hskip 0.5 cm k=(\beta
q-\gamma)+q\sqrt{(\varepsilon-\beta)(1+\frac{4\gamma}{q})}\\ \\
\left[(2\sqrt{\varepsilon-\beta}+\sqrt{1+\frac{4\gamma}{q}}~)qs-2\sqrt{\varepsilon-\beta}~\right],
& \hskip 0.5cm \mbox{for} \hskip 0.5 cm k=(\beta
q-\gamma)-q\sqrt{(\varepsilon-\beta)(1+\frac{4\gamma}{q})}\\
\end{array}\right.
\end{equation}
It is clearly seen that the energy eigenvalues are found with a
comparison of Eq.(\ref{eq7}) and Eq.(\ref{eq8}). From the four
possible forms of the polynomial $\pi (s)$ we select the one for
which the function $\tau (s)$ in Eq.(\ref{eq9}) has a negative
derivative. Therefore, the function $\tau (s)$ satisfies these
requirements, with
\begin{eqnarray}
\label{eq20}
\tau(s)=1-2qs-\left((2\sqrt{\varepsilon-\beta}+\sqrt{1+4\gamma/q}~)qs-2\sqrt{\varepsilon-\beta}~\right),\nonumber\\
\tau^{\prime}(s)=-2q-(2\sqrt{\varepsilon-\beta}+\sqrt{1+4\gamma/q}~)q.
\end{eqnarray}
Hence, the polynomial $\pi (s)$ is computed from Eq.(\ref{eq16})
as
\begin{equation}
\label{eq21}
\pi(s)=-\frac{qs}{2}-\frac{1}{2}\left[(2\sqrt{\varepsilon-\beta}+\sqrt{1+4\gamma/q}~)qs-2\sqrt{\varepsilon-\beta}~\right].
\end{equation}
From  Eq.(\ref{eq8}) we also get
\begin{equation}
\label{eq22} \lambda=(\beta
q-\gamma)-q\sqrt{(\varepsilon-\beta)(1+4\gamma/q)}-\frac{q}{2}-\frac{1}{2}\left[(2\sqrt{\varepsilon-\beta}+\sqrt{1+4\gamma/q}~)q~\right],
\end{equation}
and also
\begin{equation}
\label{eq23}
\lambda=\lambda_n=nq\left(2(1+\sqrt{\varepsilon-\beta}~)+\sqrt{1+4\gamma/q}~\right)+n(n-1)q.
\end{equation}
It is seen that the parameter $\varepsilon$ has the following form
\begin{equation}
\label{eq24} \varepsilon _{nq} =\frac{1}{16}\left[ {\sqrt
{1+\frac{4\gamma }{q}} +(1+2n)} \right]^2+\frac{\beta ^2}{\left[
{\sqrt {1+\frac{4\gamma }{q}} +(1+2n)} \right]^2}+\frac{\beta }{2}
\end{equation}
Substituting the values of $\varepsilon$ and $\beta$ into
Eq.(\ref{eq13}) and by using the transformation $2\alpha\equiv1/a$
in Eq.(\ref{eq22}), one can immediately determine the energy
eigenvalues $E_{nq}$ as
\begin{equation}
\label{eq25}
 E_{nq} =-\frac{\hbar ^2}{2ma^2}\left\{
{\frac{1}{16}\left[ {\sqrt {1+\frac{8mCa^2}{\hbar ^2q}} +(1+2n)}
\right]^2+\frac{4\left( {\frac{mV_0 a^2}{\hbar ^2}}
\right)^2}{\left[ {\sqrt {1+\frac{8mCa^2}{\hbar ^2q}} +(1+2n)}
\right]^2}+\frac{mV_0 a^2}{\hbar ^2}} \right\}.
\end{equation}
Here, the index $n$ is non-negative integers with $\infty>n\geq 0$
and the above equation indicates that we deal with a family of the
generalized Woods$-$Saxon potential. Of course it is clear that by
imposing appropriate changes in the parameters $a$ and $V_0$, the
index $n$ describes the quantization for the bound states and the
energy spectrum. In addition, if the parameter $C$ in
Eq.(\ref{eq25}) is adjusted to zero, solution reduces to the form
obtained for the standard Woods-Saxon potential without regarding
the parameter $q$.

Let us now find the corresponding wave eigenfunctions. Due to the
NU$-$method, the polynomial solutions of the hypergeometric
function $y(s)$ depend on the determination of weight function
$\rho(s)$ which is satisfies the differential equation $[\sigma
(s)\rho (s) ]^{\prime }=\tau (s)\rho (s)$. Thus, $\rho(s)$ is
calculated as
\begin{equation}
\label{eq26}
\rho(s)=\left(1-qs\right)^{\nu-1}s^{2\sqrt{\varepsilon-\beta}},
\end{equation}
where $\nu=1+\sqrt {1+\frac{4\gamma }{q}}$. Substituting into the
Rodrigues relation given in Eq.(\ref{eq4}), the eigenfunctions are
obtained in the following form
\begin{equation}
\label{eq27}
y_{nq}(s)=B_{n}\left(1-qs\right)^{-(\nu-1)}s^{-2\sqrt{\varepsilon-\beta}}\frac{d^{n}}{ds^{n}}\left[\left(1-qs\right)^{n+\nu-1}s^{n+2\sqrt{\varepsilon-\beta}}%
\right],
\end{equation}
where $B_{n}$ is the normalization constant and its value is
$1/n!$. Choosing $q=1$, the polynomial solutions of $y_n(s)$ are
expressed in terms of Jacobi Polynomials, which is one of the
orthogonal polynomials with weight function
$(1-s)^{\nu-1}s^{2\sqrt{\varepsilon-\beta}}$ in the closed
interval $[0,1]$, giving
[constant]$P_n^{(2\sqrt{\varepsilon-\beta},~\nu-1)}(1-2s)$ \cite
{27}. By substituting $\pi(s)$ and $\sigma(s)$ into the expression
$\phi (s)^{\prime }/\phi (s)=\pi (s)/\sigma (s)$ and solving the
resulting differential equation, the other part of the wave
function in Eq.(\ref{eq2}) is found as
\begin{equation}
\label{eq28} \phi(s)=(1-s)^{\mu}s^{\sqrt{\varepsilon-\beta}},
\end{equation}
where $\mu=\nu/2$ and again taking $q=1$. Combining the Jacobi
polynomials and $\phi(s)$ in Eq.(\ref{eq28}), the s$-$wave
functions are constructed as
\begin{equation}
\label{eq29}
\psi_n(s)=A_ns^{\sqrt{\varepsilon-\beta}}(1-s)^{\mu-\nu+1}P_n^{(2\sqrt{\varepsilon-\beta},~\nu-1)}(1-2s),
\end{equation}
where $A_n$ is a new normalization constant.

\section{Potential and Energy Eigenvalue Shapes}
For the generalized Woods-Saxon potential in Eq.(\ref{eq10}), the
potential shape can be determined by the parameters $V_0$, $a$ and
$C$. Therefore, we have adopted a generalized Woods$-$Saxon shape
for the real part of the optical$-$model potential \cite {29} and
investigated the scattering phenomenon for this potential only.
The empirical values found by Perey et al. \cite {30}
$r_0=1.285~fm$, $a=0.65~fm$ and $V_0\approx40.5+13A~MeV$ are used.
Here, $A$ is the mass number of target nucleus and
$R_0=r_0A^{1/3}$. The shape of the generalized Woods$-$Saxon
potential given by Eq.(10) is illustrated in Fig.$1$.\\

Fig.2 shows that the energy eigenvalues as a function of the
discrete level $n$ for different values of the parameter $a$. Of
course it is clear that by imposing appropriate changes in the
parameters $a$ and $V_0$, the index $n$ describes the quantization
of the bound states and the energy spectrum. Some of the initial
energy levels for $q=1$ value are presented by choosing
$C=10~MeV$, $A=56$ which is the geometric average of the target
mass number $44\leq A \leq72$ \cite {31}.

\section{Conclusions}
The exact solutions of the radial Schr\"odinger equation for the
generalized Woods-Saxon potential with the zero angular momentum
are obtained. Eigenvalues and eigenfunctions obtained from the
real form of the potential are computed. So, the wave functions
are physical and energy eigenvalues are in a good agreement with
the results obtained by the other methods. Nikiforov-Uvarov method
is used in the calculations. We have seen that there were some
restriction on the potential parameters for the bound states
within the framework of quantum mechanics. That is, the value of
$C$ from the potential parameters is increased for the constant
value of the parameter $a$, it is seen that the depth of potential
increases rapidly. Therefore, if all the parameters of potential
remain purely real, it is clear that all bound energies $E_n$ with
$n\geq 0$ represent a negative energy spectrum. We also point out
that the exact results obtained for the generalized Woods-Saxon
potential may have some interesting applications in the study of
different quantum mechanical and the nuclear scattering systems.

\newpage

\vspace{0.5in} \noindent{\Large \bf Figure Captions} \vskip .5
true cm

\noindent {\bf Figure 1:} Variation of the generalized Woods-Saxon
potential as a function of r.\\

\noindent {\bf Figure 2:} The variation of the energy eigenvalues
with respect to the quantum number $n$ with $V_0\approx 48$  MeV.
The curves correspond to the different values of the range parameter $a$.\\

\newpage

\begin{figure}
\epsfig{file=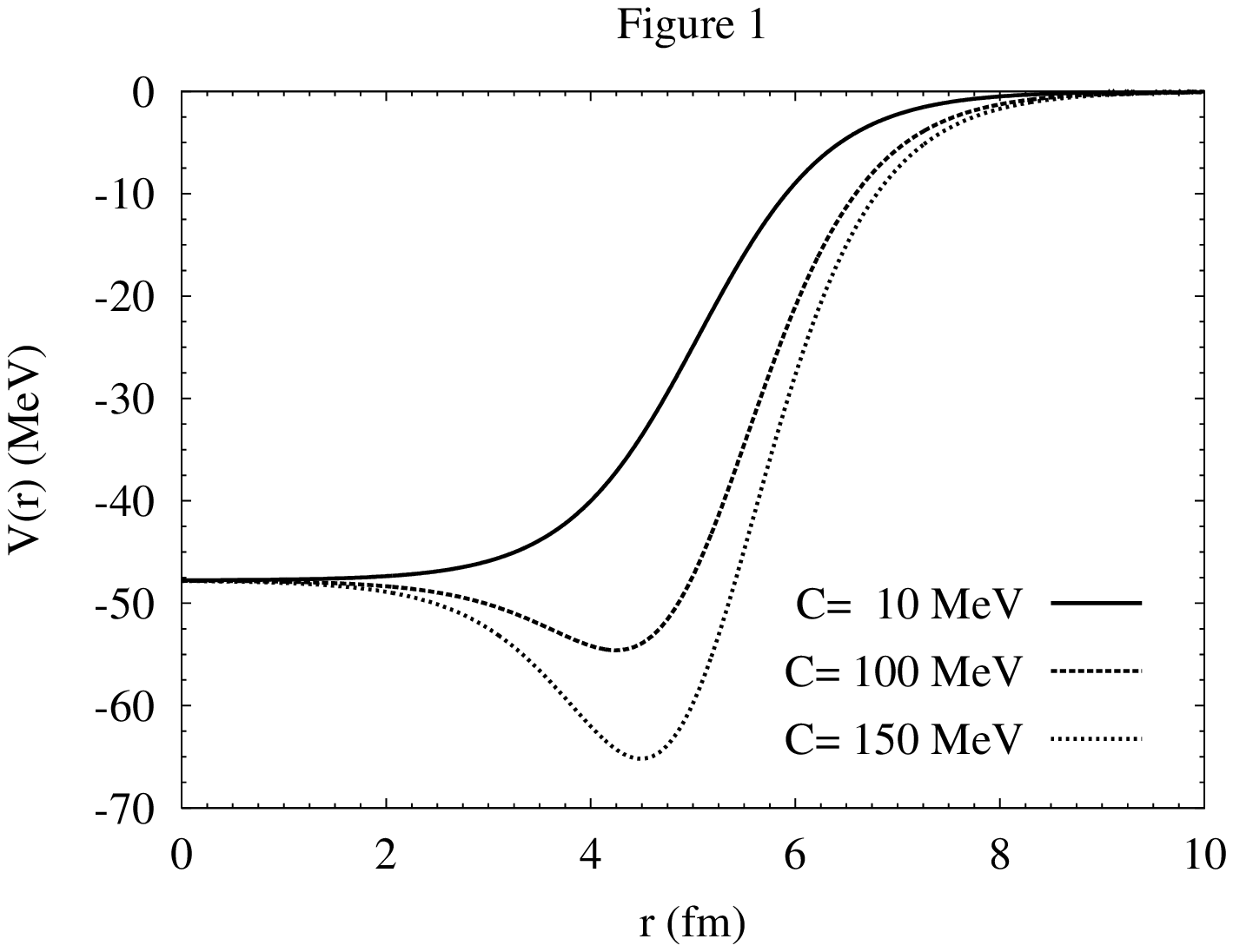,width=17cm,angle=0} \label{Fig1}
\end{figure}

\newpage

\begin{figure}
\epsfig{file=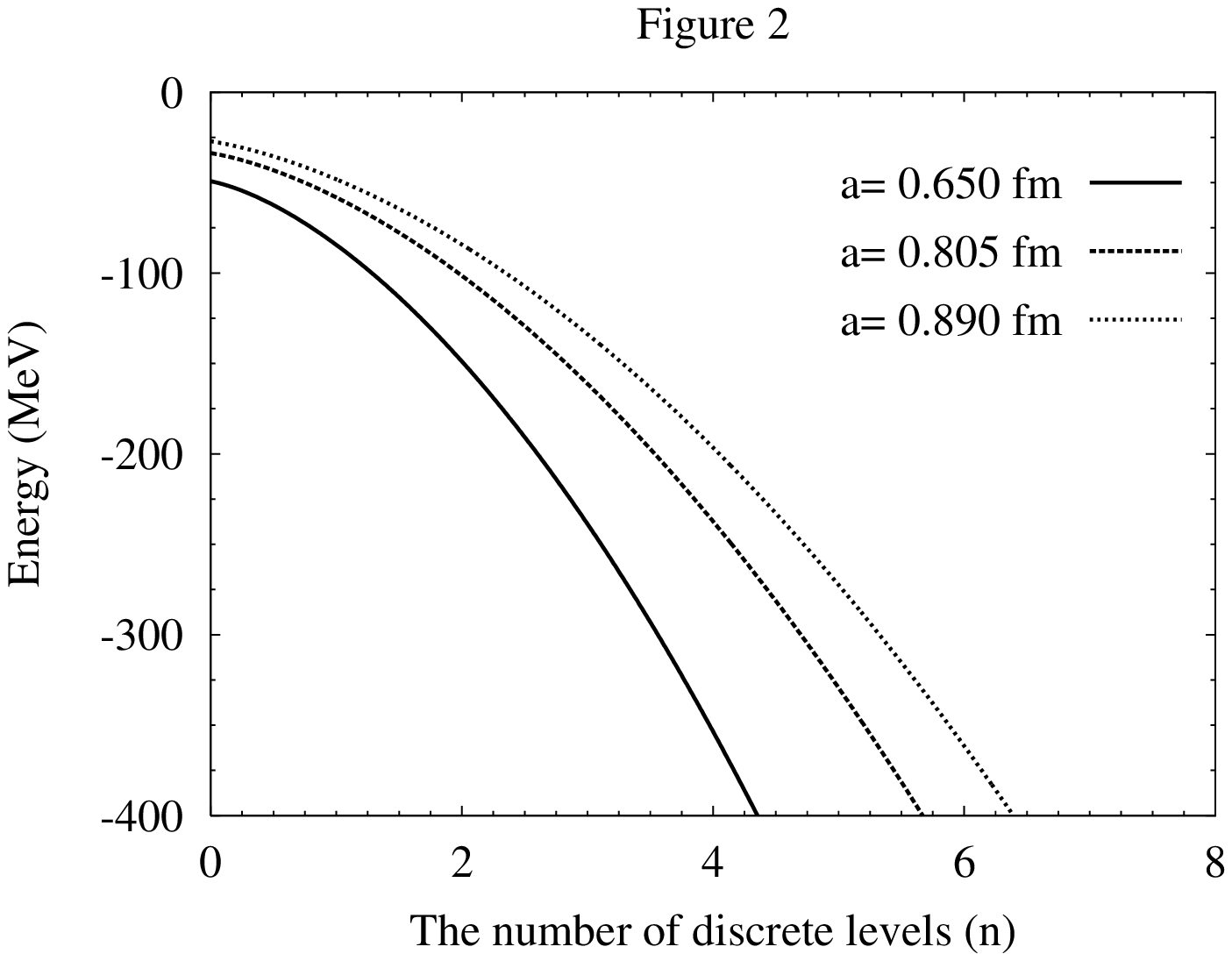,width=17cm,angle=0}
\label{Fig2}
\end{figure}

\end{document}